\documentclass[showpacs,twocolumn,aps,prl]{revtex4}
\usepackage{amsmath}
\usepackage{ulem}
\usepackage{amsfonts}
\usepackage{amssymb}
\usepackage{amsthm}
\usepackage{graphicx}
\usepackage{CJK}
\usepackage{subfigure}
\begin{document}

\title{Instability of Walker Propagating Domain Wall
in Magnetic Nanowires}
\author{B. Hu}
\author{X.R. Wang}
\email{[Corresponding author:]phxwan@ust.hk}
\affiliation{Physics Department, The Hong Kong University of
Science and Technology, Clear Water Bay, Kowloon, Hong Kong}
\date{\today}
\begin{abstract}
Stability of the well-known Walker propagating domain wall (DW)
solution of the Landau-Lifshitz-Gilbert equation is analytically
investigated. Surprisingly, the Walker's rigid body propagating
DW mode is not stable against the spin wave/wavepacket emission.
In the low field region only stern spin waves are emitted while
both stern and bow waves are generated under high fields.
In a high enough field, but below the Walker breakdown field,
the Walker solution could be convective/absolute unstable if the
transverse magnetic anisotropy is larger than a critical value,
corresponding to a significant modification of the DW profile and
DW propagating speed.
\end{abstract}
\maketitle

Magnetic domain-wall (DW) propagation in nanowires has attracted
considerable attention in recent years \cite{Walker,Parkin,
Cowburn,Erskine,Wang} because of its fundamental interest and
potential applications \cite{Parkin,Cowburn}. Field-driven DW
dynamics is governed by the Landau-Lifshitz-Gilbert (LLG) equation
which has a well-known Walker's exact rigid-body propagation
solution \cite{Walker} for a one-dimensional (1D) biaxial wire.
This Walker solution plays a pivotal role \cite{Zhang,Fert,Linder}
in our current understanding of both current-driven and field-driven
DW propagation in magnetic nanowires. A genuine solution
of a physical system must be stable against small perturbations.
Although there is no proof of the stability of the Walker solution
and there are signs \cite{Wieser,Xiansi} that this solution may
be unstable, at least under certain conditions, the validity of
the Walker solution for a 1D wire is always taken as self-evident.
Any deviation in experiments or numerical simulations are assumed to
be attributed to the quasi-1D nature or other effects \cite{Fert}.
On the other hand, applications of spintronics devices require
accurate description of DW motion \cite{Stohr,Gerrit1,Han,magnon}.
Thus, the stability of the Walker propagating DW solution becomes
vital in our understanding of DW propagation along a magnetic wire.

In this paper, by using a recipe that is based on a series of
recent advances in nonlinear dynamics theory, the stability of
the Walker's exact DW solution is theoretically analyzed.
To our surprise, the solution is not stable against spin wave emission.
In the low field region, only stern spin wave could be observed
while both stern and bow waves emerge under high field.
Severe distortion in propagating DW profile and resulted deviation
of DW speed from the Walker formula can occur when the transverse
magnetic anisotropy is larger than a critical value and the external
field is sufficient high, but below the Walker breakdown field.
For a given transverse magnetic anisotropy, the solution is
transient unstable at low field and convective/absolute unstable at
high field, corresponding to emitted different spin wavepackets.
\begin{figure}
  \includegraphics[width=3.4 in]{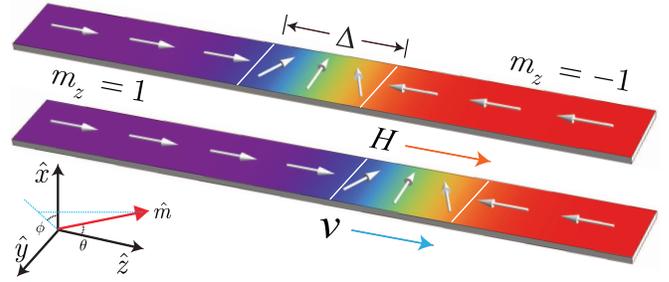}\\
\caption{(Color online) Illustration of transverse head-to-head DW
of width $\Delta$ in a nanowire, with easy axis along $\hat z$ and
hard axis along $\hat x$. In the absence of external magnetic field
(upper), a static DW exists between two domains with $m_z=\pm 1$.
Under a field parallel to the easy axis, the Walker propagating DW
moves towards the energy minimum state ($m_z=-1$) at a speed $v$
while the DW profile is preserved. }
\label{fig1}
\end{figure}

To study the stability of Walker's exact propagating DW
solution under an external field, we consider the dimensionless
1D LLG equation \cite{Wang},
\begin{equation}
\frac{\partial \vec m}{\partial t} = -\vec m\times{\vec h_{eff}}
+ \alpha \vec m \times \frac{{\partial \vec m}}{{\partial t}}.
\label{llg} \end{equation}
This LLG equation describes the dynamics of magnetization $\vec M$
of a magnetic nanowire schematically shown in Fig. \ref{fig1}.
With the easy axis along the wire ($\hat z$ direction) and the width
and thickness being smaller than the exchange interaction length,
exchange interaction dominates the stray field energy caused by magnetic
charges on the edges; the DW structure tends to be homogeneous in the
transverse direction \cite{Porter}, i.e., behaves effectively 1D.
We are interested in the behavior of a head-to-head DW under an
external field shown in Fig. \ref{fig1}. In Eq. \eqref{llg}, $\vec m$
is the unit direction of the local magnetization $\vec M= \vec m M_s$
with saturation magnetization $M_s$ and $\alpha$ is the
phenomenological Gilbert damping constant. The effective field (in the
units of $M_s$) is $\vec h_{eff}=K_\parallel m_z \hat{z}-K_\perp m_x
\hat{x}+A \partial^2\vec m/\partial z^2 +H\hat{z}$ where $K_\parallel$,
$K_\perp$, and $A$ are respectively the easy axis anisotropy coefficient,
the hard axis anisotropy coefficient, and the exchange coefficient.
$H$ is the external magnetic field parallel to $\hat{z}$.
The time unit is $(\gamma M_s)^{-1}$, where $\gamma$ is the gyromagnetic
ratio. Using polar angle $\theta$ and azimuthal angle $\varphi$ for
$\vec m$ as shown in Fig. \ref{fig1}, this LLG equation has a well
known Walker propagating DW solution \cite{Walker},
\begin{equation}
 \begin{split}
\sin 2{{\varphi_w(z,t)}} =\frac{H}{H_c}, \quad
\ln \tan \frac{1}{2}{\theta_w(z,t)}  =  \frac{z-vt}{\Delta}.
\end{split}
\label{walker} \end{equation}
Here $H_c=\alpha K_\perp /2$ is the Walker breakdown field and $\Delta=
{(K_\parallel/A + {\cos^2}{\varphi_w}K_\perp/A)^{-1/2}}$ is the DW width
which will be used as the length unit ($\Delta=1$) in the analysis below.
$v=\Delta H/\alpha$ is the Walker rigid-body DW speed that is linear
in the external field and the DW width, and inversely proportional to the
Gilbert damping constant. Solution \eqref{walker} is exact for $H<H_c$.
\begin{figure}
  \includegraphics[width=3.4 in]{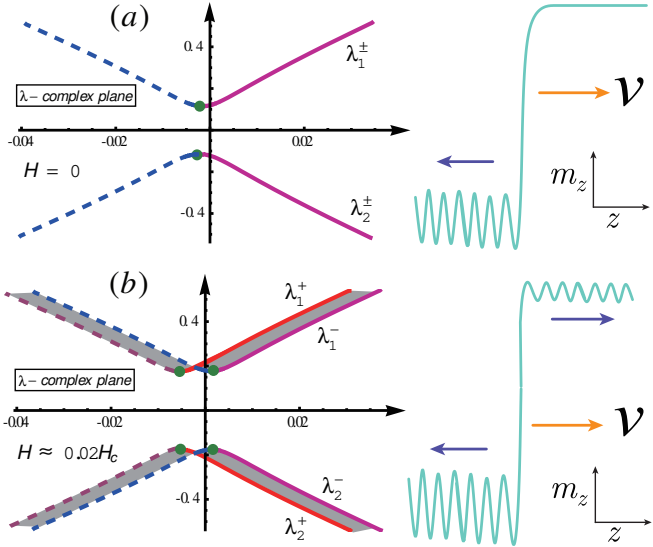}\\
\caption{(Color online) Left are the essential spectrum (shadowed regions)
for $H=0$ (a) and \textbf{$H \approx 0.02 H_c$} (b). The Fredholm borders are
$\lambda_{1,2}^{\pm}(k)$. Solid border lines correspond to spin waves
with negative group velocities  while the dashed border lines
are for the spin waves with positive group velocities.
Propagating DW wall emit stern waves in low fields (right of (a)), and
stern and bow waves in higher field ($0.02H_c<H<H_c$) (right of (b)).
The green dots are zero group velocity modes. $K_\perp=1$ is used. }
\label{fig2}
\end{figure}

To prove the instability of solution \eqref{walker} against spin
wave emission, we follow a recently developed theory (Sandstede, 2001
\cite{Sandstede}; Fiedler and Scheel, 2003 \cite{Fiedler})
for stability of a general travelling front such as
a propagating head-to-head DW shown in Fig. \ref{fig1}.
Consider a small deviation of the Walker solution, ${\theta_w}+\theta$
and ${\varphi_w}+ \varphi$ with $|\theta|\ll 1$ and $|\varphi|\ll 1$, the
equations satisfied by $\theta$ and $\varphi$ can be readily obtained
from Eq. \eqref{llg}. In the moving frame of the DW velocity $v$
(with coordinate transformations of $z \rightarrow \xi\equiv z-vt$ and
$t \rightarrow t$), the linearized equations of $\theta$ and $\varphi$
in a two-component form of $\Lambda  \equiv {(\theta,\;\varphi)^T}$
(superscript T means transpose) are
  \begin{equation}
\frac{{d \Lambda }}{{dt}} = L({\theta_w},{\varphi_w},{\partial }/{{\partial
\xi }},{{{\partial ^2}}}/{{{\partial ^2}\xi }})\Lambda.
\label{linearize}
\end{equation}
$L$ is an inhomogeneous operator, depending on $\xi$ through $\theta_w$.
The energy dissipation due to time variations of $\theta$ and $\varphi$
are absent because they are higher order terms \cite{Wang}.
It shall be important when the deviations are out of the linear regime.
The possible solutions of Eq. \eqref{linearize} of type $\Lambda=
\Lambda_1(\xi)e^{\lambda t}$ define spectrum $\Lambda$ of $L$.
Similar to energy spectrum of a quantum system, $\lambda$ can be
continuum and discrete. The former is often called essential spectrum
while the later point spectrum. The spectrum $\lambda$ shall determine
the stability of the Walker solution. In terms of
$\Lambda '\equiv {(\theta, \;\varphi,\;\partial \theta /\partial
\xi ,\;\partial \varphi /\partial \xi )^T}$, equation
\eqref{linearize} becomes a four dimensional first-order ordinary
differential system
\begin{equation}
\frac{d}{{d\xi }}\Lambda ' = \Gamma \Lambda ',\quad
\Gamma  = \left( {\begin{array}{*{20}{c}}0&I,\\
B&C \end{array}} \right),
\label{linearize2}
\end{equation}
here $I$ is a $2\times 2$ identity matrix, and $2\times 2$ matrices,
$B$ and $C$, have following matrix elements:
$B_{11} =\alpha\lambda/A +(1/2A)(K_\bot-2K_\parallel - K_\bot
\sqrt{1 -\rho^2})\cos [2G(\xi)] - H\tanh \xi/A$ where $G(\xi)$ is
the Gudermannian function and $\rho=H/H_c$;
$B_{12} = (\lambda  - K_\bot \rho \tanh \xi)/(A\cosh \xi) $;
$B_{21} = -\cosh \xi(2\lambda + K_\bot \rho \tanh \xi)/(2A)$;
$B_{22} = (\alpha\lambda -K_\bot \sqrt {1 -\rho^2})/A$;
$C_{11} = - v\alpha/A $; $C_{12}=- v/(A\cosh \xi)$;
$C_{21}=v\cosh \xi/A$; $C_{22} =  -2 - v\alpha/A$.

According to the theory of Refs. \cite{Sandstede,Palmer,Henry,Fiedler,
nbook}, the spectrum of $L$ is determined by Eq. \eqref{linearize2}.
The essential spectrum is bordered by the Fredholm borders (defined
below) of \eqref{linearize2} with $\Gamma$ replaced by its
two limits of $\xi \rightarrow \pm \infty$, denoted
as $\Gamma^{\pm}\equiv{\lim _{\xi  \to  \pm \infty }}\Gamma $.
Since $\Gamma^{\pm}$ are constant $4\times 4$ matrices,
solutions of Eq. \eqref{linearize2} with $\Gamma=\Gamma^{\pm}$
are linear combinations of $\Lambda_0 e^{\kappa^{\pm} \xi} $ with
$\kappa^{\pm}$ being complex numbers.
Pure plane wave solutions ($\kappa^{\pm}=ik$) exist only when $\lambda$
satisfies $\det(\Gamma^{\pm}(\lambda)+ik)=0$  with $k\in(-\infty,\infty)$.
Each of the two equations has two branches of allowed $\lambda$ labeled
as $\lambda_{1,2}^{\pm}(k)$, known as the Fredholm borders
\cite{Sandstede,Palmer,Fiedler}.
In another word, Eq. \eqref{linearize2} with $\Gamma=\Gamma^+$
($\Gamma=\Gamma^-$) has pure plane wave solution when $\lambda$ is on
$\lambda_{1,2}^+(k)$ ($\lambda_{1,2}^-(k)$). For those $\lambda$ not on
$\lambda_{1,2}^{\pm}(k)$, each Eq. \eqref{linearize2} with $\Gamma
=\Gamma^{\pm}$ have four $\kappa^{\pm}$'s whose real parts are nonzero.
If one uses $(n^{\pm}_+,n^{\pm}_-)$ to denotes $\lambda$ for $n^{\pm}_+$
($n^{\pm}_-$) being the number of $\kappa^{\pm}$'s with positive (negative)
real part. Then both $\lambda_{1,2}^+(k)$ and $\lambda_{1,2}^-(k)$ divide
$\lambda$-plane into three parts with $(n^{\pm}_+,n^{\pm}_-)=(1,3),\ (2,2),$
and $(3,1)$, respectively. According to theorem \textbf{5.35} in Chapter 4
of Kato \cite{nbook}, the essential spectrum of $L$ must be in the regimes
with $n^+_-+n^{-}_{+}\neq 4$. For $\lambda$ on boundaries $\lambda_{1,2}^\pm$,
the associated eigenmode is plane wave (spin wave) while eigenmodes for
$\lambda$ not on the boundaries are spin wavepackets.

\begin{figure}
  \includegraphics[width=3.4 in]{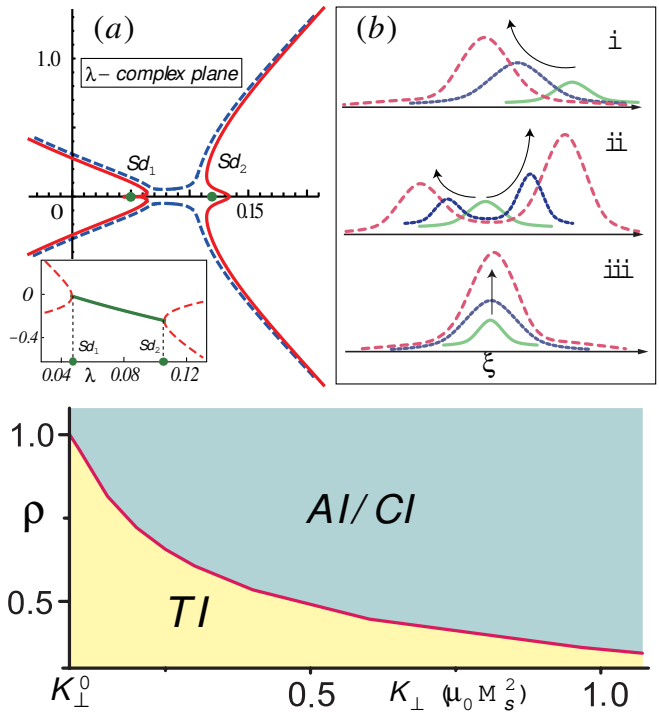}\\
\caption{(Color online)
(a) $\lambda_{1,2}^-$ for $K_\perp=1$, $\rho=0.35$ (dashed curve) and
$0.36$ (solid curve). The absolute spectrum is between two branching
points $Sd_1$ and $Sd_2$ (green dots).
Inset: Plot of $Re (\kappa^-_2)$ and $Re (\kappa^-_3)$ vs. $\lambda$
between $Sd_1$ and $Sd_2$. At $Sd_{1,2}$, $\kappa^-_2=\kappa^-_3$.
(b) Graphical illustrations of three types of instabilities caused by
unstable absolute spectrum. Green curves indicate
initial profiles of unstable modes while the dotted (blue) and dashed
(red) curves are their later profiles. A transient instability (i) emits
unidirectional waves (propagating to the left).
A convective instability (ii) emits waves in both directions.
An absolute instability (iii) emits waves that do not travel in the
moving frame, or move with the DW.
(c) Phase diagram of transient (TI) and
absolute/convective (AI/CI) instabilities. The boundary is the bifurcation
line between TI/CI-and-CI instabilities in $K_\perp$ and $\rho=H/H_c$ plane.
The bifurcation line is only plotted for
$K_\perp \geq K_\perp^0 $ here $K_\perp^0\approx 0.085$ at which $H_2=H_c$
($\rho=1$). Noted that our analysis is valid for fields below the Walker
breakdown value.}
\label{fig3}
\end{figure}

In order to understand numerical results in Ref. \cite{Xiansi}, parameters
of yttrium iron garnet (YIG) \cite{magnon} are assumed in our analysis with
$A= 3.84 \times 10^{-12}  J/m$, $K_{\parallel}=2 \times 10^3 J/m^3$,
$\gamma=35.1 kHz/(A/m)$, and $M_s=1.94 \times 10^5 A/m$.
$\alpha=0.001$ is used and $K_{\perp}$ is a varying parameter.
Fig. \ref{fig2} plots the essential spectrum for $K_\perp=1$.
In the absence of an external field, two branches of the spectrum of
$\Gamma^\pm$ are the same, $\lambda_{1,2}^{+}(k)=\lambda_{1,2}^{-}(k)$,
shown in Fig. \ref{fig2}(a). Since the spectrum encroaches the right half
plane, unstable plane waves shall exist and spin wave emission are expected.
Similar conclusion was also obtained in early study \cite{Bouzidi}.
Solid lines are for negative group velocity (determined by
$Im ({\partial \lambda}/{\partial k})$), thus these are stern modes.
The dashed lines indicate positive group velocity, corresponding to
bow modes. The green dots are zero group velocity points.
According to Fig. \ref{fig2}(a), all unstable modes have negative group
velocities so that DW can only emit stern waves in the low fields.
As the external field increases, $\lambda_{1,2}^{+}(k)$ and $\lambda_{1,2}
^{-}(k)$ are different, and the area of essential spectrum in $\lambda
$-plane becomes bigger and bigger (shadowed regimes in Fig. \ref{fig2}(b).
The green dots also moves toward $Im(\lambda)$-axis and cross it at
$H \approx 0.02H_c$ (Fig. \ref{fig2}(b)).
Further increase of $H$, the unstable modes have both positive and
negative group velocities although the most of them have the negative ones.
One shall have propagating DW to emit both stern and bow waves.
The stern waves should be stronger than the bow waves as schematically
shown in the right figure of Fig. \ref{fig2}(b). This is exactly what
were observed in numerical simulations on dissipative wires for stern
wave emission in low field \cite{Wieser} and stern-and-bow wave
emission in high field \cite{Xiansi}. In a realistic wire with damping,
emitted spin waves will be dissipated after a short distance, and are
hard to be observed in experiments.

Interestingly, severe instabilities of a propagating DW are not
determined by the essential spectrum, but by the absolute spectrum
\cite{Sandstede,Palmer,Henry,Fiedler,nbook,Sandstede2,Chomaz,Brevdo}
that could change the DW profile so that DW propagating speed would be
substantially modified \cite{Wang}. As mentioned early, for each $\lambda$
in the complex plane, there are four $\kappa^\pm_i$ ($i=1,2,3,4$) for
$\Gamma^\pm$, ordered by their real parts as $Re(\kappa^\pm_1)
\geq Re(\kappa^\pm_2)\geq Re(\kappa^\pm_3)\geq Re(\kappa^\pm_4)$.
Then $\lambda$ is said to belong to the absolute spectrum if and only
if $Re (\kappa^+_2(\lambda)) = Re (\kappa^+_3(\lambda))$ or $Re (\kappa
^-_2(\lambda)) = Re (\kappa^-_3(\lambda))$ \cite{Sandstede2,Rademacher}.
The branching points are special points in the absolute spectrum, denoted
as $\lambda_{sd}$, satisfying $\kappa^\pm_2(\lambda_{sd}) =\kappa^\pm_3
(\lambda_{sd})$. They are non-traveling modes \cite{Sandstede2,Rademacher}.
For $K_\perp=1$, the absolute spectrum in the right half $\lambda$-plane
is generated by $\Gamma^-$. Fig. \ref{fig3}(a) shows two branches
$\lambda_{1,2}^-$. They are well separated by the real axis for
$\rho=0.35$ as shown in Fig. \ref{fig3}(a) (dashed curves) and
no absolute spectrum could be found in the right half plane.
As the field increases, the two branches get closer with each other and
at an onset field $H_2$, depending on  $K_\perp$, two branches tangent
at the real axis and then separate again in horizontal direction
as shown in Fig. \ref{fig3}(a) for $\rho=0.36$ (solid curves).
At this moment, the absolute spectrum begin to emerge on the real axis
(the segment between two branching points $Sd_{1,2}$ (green solid dots).
The dependence of $Re (\kappa_2^-)$ or $Re (\kappa_3^-)$ on
$Re(\lambda)$ between these two points is shown in the inset of Fig.
\ref{fig3}(a) (solid segment).

According to Refs. \cite{Sandstede2,Chomaz,Brevdo}, wavepackets
would be emitted if the essential spectrum encroaches the right
half $\lambda$-plane. There are three types of instability
\cite{Sandstede,Palmer,Fiedler,Henry,Sandstede2,Chomaz,Brevdo}.
The instability is called transient (TI) if the essential
spectrum encroaches the right half plane and absolute spectrum are
either in the left half plane or does not exist. The propagating DW
emits stern waves shown in Fig. \ref{fig3}(b)i. The instability is
called convective if both essential and absolute spectrum encroaches
the right half $\lambda$-plane. In this case, the emitted waves
can propagate in both direction as shown by Fig. \ref{fig3}(b)ii.
For an convective instability, if any branching point is also in
the right half $\lambda$-plane, the instability is called absolute.
An absolute instability can then emit non-traveling (zero group
velocity) waves as illustrated in Fig. \ref{fig3}(b)iii.
For LLG equation, since the absolute spectrum is the segment
connecting two branching points $Sd_{1}$ and $Sd_{2}$ (Fig. \ref{fig3}
(a)), the absolute instability (AI) and convective instability (CI)
co-exist. It is known that transient instability is very weak that
can be removed under proper mathematical treatment
\cite{Sandstede,Humpherys}. Thus, we should not expect to have great
physical consequences.
On the other hand, the absolute instability move with the DW, and
cause the change of DW profile \cite{Pego,Sandstede2,Humpherys}.
It is known \cite{Wang} that field-induced DW propagating speed is
proportional to the energy damping rate that is sensitive to DW profile.
Therefore absolute instability, which deform propagating DW profile,
shall substantially alter DW speed. This may explain why the
field-induced DW speed start to deviate from the Walker result
only when the field is large enough to emit both stern and bow
waves in simulations \cite{Xiansi}.

Fig. \ref{fig3}(c) is the calculated phase diagram in
$K_\perp$ and $\rho=H/H_c$ plane. A transition from transient
instability (denoted as TI in the figure) to absolute/convective
instability (AI/CI) occur at a critical field $H_2$ as lng as
$K_\perp>K_\perp^0 \approx 0.085$ at which $H_2=H_c$.
It means no absolute/convective instability exist for
$K_\perp<K_\perp^0$, and one shall not see noticeable change in
famous Walker propagation speed mentioned early. This may
explain why many previous numerical simulations on permalloy,
which have small transverse magnetic anisotropy, are consistent
with Walker formula.
A snapshot of the convecting wavepackets could be identified in
Fig. 2 in Reference \cite{Xiansi} where wavepackets can be seen
in the vicinity of the traveling DW and travel to
both directions.

It should be noticed that the effects of point spectrum have not
been analyzed. In principle, it can also affect the stability of
the Walker solution, and should be a very interesting subject too.
Unfortunately, there are not many theorems on the point spectrum
yet. Thus, one can only rely on a numerical method to find a
point spectrum of operator $L$ and to find out whether it can
also induce any instability on a propagating DW.

In conclusion, we showed that a Walker propagating DW will
always emit stern waves in a low field, and both stern and
bow waves in a higher field. Thus the exact Walker solution
of LLG equation is not stable. The true propagating DW
is always dressed with spin waves. The emitted spin waves
shall be damped away during their propagation, and make
them hard to be detected in realistic wires.
For a realistic wire with its transverse magnetic anisotropy
larger than a critical value and when the applied external field
is larger than certain value, a propagating DW may undergo
simultaneous convective and absolute instabilities.
As a consequence, the propagating DW will not only emit
both spin waves and spin wavepackets, but also change
significantly its profile.
Thus, the corresponding Walker DW propagating speed will
deviate from its predicted value, agreeing very well with
recent simulations.

This work is supported by Hong Kong RGC Grants (604109 and RPC11SC05).

\end{document}